\documentclass[aps,showpacs,amsfonts,amsmath,prb,endfloats*,preprint]{revtex4}

\usepackage{graphicx}  

\usepackage{color}

\begin{document}
\title{Micelle formation in block copolymer/homopolymer blends: comparison of self-consistent field theory with experiment and scaling theory}
\date\today
\author{Martin J.~Greenall}
\affiliation{School of Physics and Astronomy, University of Leeds,
Leeds LS2 9JT, U.K.}
\author{D.~Martin A.~Buzza}
\affiliation{Department of Physics, The University of Hull,
Cottingham Road, Hull HU6 7RX, U.K.}
\author{Thomas C.~B.~McLeish}
\affiliation{School of Physics and Astronomy, University of Leeds,
Leeds LS2 9JT, U.K. and Department of Physics, Durham
University, South Road, Durham DH1 3LE, U.K.}

\begin{abstract}
We present a self-consistent field theory (SCFT) study of
spherical micelle formation in a blend of poly(styrene-butadiene)
diblocks and homopolystyrene. The micelle core radii, corona
thicknesses and critical micelle concentrations are calculated as
functions of the polymer molecular weights and the composition of
the diblocks. We then make a parameter free comparison of our
results with an earlier scaling theory and X-ray scattering data.
For the micelle core radii $R_c$, we find that SCFT reproduces the
shape of the variation of $R_c$ with different molecular parameters
much more accurately compared to scaling theory, though like scaling
theory, it overestimates $R_c$ by about 20-30\%. For the corona
thickness $L_c$, the accuracy of our SCFT results is at least as
good as those of scaling theory. For copolymers with lighter core
blocks, SCFT predictions for the critical micelle concentration
improve over those of scaling theories by an order of magnitude. In
the case of heavier core blocks however, SCFT predicts the critical
micelle concentration less well due to inaccuracies in the modeling
of the bulk chemical potential. Overall, we find that SCFT gives a
good description of spherical micelle formation and is generally
more successful than scaling theory.
\end{abstract}
\pacs{36.20.Ey, 47.57.Ng, 61.25.he, 64.75.Va, 64.75.Yz}

\maketitle

\section{Introduction}

When dissolved in a solvent such as water, amphiphilic block
copolymers can self-assemble into micelles similar to those seen in
low molecular weight surfactants \cite{jain_bates}. The {\em
solvophobic} blocks cluster together to form a core, and the {\em
solvophilic} blocks spread outwards as a corona. A certain minimum
concentration is needed for micelles to form: this is the {\em
critical micelle concentration}, or cmc \cite{safran_book}. These
block copolymer micelles have been the subject of growing recent
interest, from the points of view of both fundamental polymer
physics \cite{zhulina} and potential applications, particularly in
drug delivery \cite{kim}.

A closely related system, which is the subject of this paper, is a
blend of block copolymer and homopolymer. Here, the block copolymers
may also form micelles, with the homopolymer acting as the solvent
\cite{kinning, roe}. This system is more easily controlled and
better characterized than an aqueous solution of diblocks, and so is
more suitable for a study of the fundamentals of micelle formation.
In particular, the interaction strengths between the different
components are well described by the standard $\chi$ parameter
\cite{matsen_book}, and can be obtained from the literature
\cite{polymer_handbook}.

Detailed experimental data are available on spherical micelle
formation in these blends \cite{rigby_roe1, rigby_roe, kinning}.
These data have been modeled using scaling theories \cite{leibler,
whitmore_noolandi, mayes_delacruz, roe, kinning}. However, whilst
qualitatively reproducing the general trends seen in these systems
(for example, the increase in the micelle core radius as the
homopolymer length is increased), scaling often fails to predict
the shape or magnitude of these variations quantitatively. In this
paper, we use a more microscopic approach which involves less severe
approximations: {\em self-consistent field theory} (SCFT)
\cite{edwards, matsen_book}. SCFT, a mean-field theory of an
ensemble of flexible polymers, is probably the most successful
theoretical method for the modeling of the structures formed by
polymers, and provides quantitative results on polymer melts
\cite{matsen_book}. To our knowledge, the properties of micelles in
copolymer/homopolymer blends have not been studied in detail, using
experimentally-determined polymer properties, with SCFT. Although a
considerable body of work exists on the self-consistent field theory
of micelle formation, this mostly focuses on diblocks dispersed in
{\em monomer} solvent
\cite{leermakers,leermakers_scheutjens-shape,linse} or on diblock
copolymer melts \cite{wang_wang_yang} (although see the paper by
Duque on intermediate states in micelle formation \cite{duque} and
that by Monzen {\em et al} \cite{monzen} on {\em triblocks} in
homopolymer). A further motivation for the study of micelles using
SCFT is that much previous research using this approach has
concentrated on periodic structures \cite{matsen_book}. In contrast
to this, our work is a detailed test of the theory for isolated
aggregates, and provides an assessment of the suitability of SCFT
for the investigation of the self-assembly of polymers in solution
and the design of macromolecules for applications.

We focus on the X-ray scattering experiments of Kinning, Thomas and
Fetters \cite{kinning}, which study poly(styrene-butadiene) diblocks
in homopolystyrene. We compare our predictions with their results on
the core and corona radii and the critical micelle concentration.
Our SCFT results are also compared with the scaling theory
predictions of Leibler, Orland, Wheeler and others \cite{leibler,
whitmore_noolandi, mayes_delacruz, roe}.

\section{Details of system and scaling theory}\label{system_details}

Kinning, Thomas and Fetters \cite{kinning} studied blends of
poly(styrene-butadiene) diblock copolymer and polystyrene
homopolymer. In order to determine the effects of the molecular
weight of the polymers and the relative amounts of styrene and
butadiene in the diblocks, several samples were considered. We adopt
the notation used by these authors \cite{kinning} to label the
samples. As an example, a diblock sample of polystyrene molar weight
60 kg/mol and polybutadiene molar weight 10 kg/mol is referred to as
SB 60/10. A polystyrene homopolymer sample of molar weight 2.1
kg/mol is labeled 2100 PS. Note that these values are quite rough:
more precise molecular weights for the individual samples are
tabulated in the original paper \cite{kinning}.

The polymer samples are characterized by their specific volumes, the
root mean square end-to-end distances of the polymer molecules, and
the polystyrene-polybutadiene {\em interaction energy density}. We
now introduce and discuss these quantities.

The specific volume of PB as a function of temperature is given
empirically by \cite{polymer_handbook}
\begin{equation}
V_\text{PB}(\text{cm}^3/\text{g})\approx 1.0968 + (8.24\times
10^{-4})T \label{V_PB}
\end{equation}
while that of PS has been found \cite{richardson_savill} to be
\begin{equation}
V_\text{PS}(\text{cm}^3/\text{g})\approx 0.9217 + (5.412\times
10^{-4})T + (1.678\times 10^{-7})T^2 \label{V_PS}
\end{equation}
The temperature $T$ is measured in degrees Celsius, and is equal to
$115^\circ\text{C}$ in all experiments considered here.

SCFT is more naturally written in terms of the volumes of individual
molecules $v_i$, with $i$ representing either PB or PS. These can be
calculated from Equations \ref{V_PB} and \ref{V_PS} by
\cite{kinning}
\begin{equation}
v_i({\text{\AA}^3})=M_i(\text{g/mol})\times
V_i(\text{cm}^3/\text{g})/0.602 \label{molecular_volume}
\end{equation}
where $M_i$ is the molar weight of substance $i$.

The polymer root-mean-square end-to-end distances are given, also
empirically, by \cite{kinning}
\begin{eqnarray}
\langle R^2_\text{PB}\rangle^{1/2}(\text{\AA}) & \approx & 0.93M_\text{PB}^{1/2}\nonumber\\
\langle R^2_\text{PS}\rangle^{1/2}(\text{\AA}) & \approx &
0.70M_\text{PS}^{1/2} \label{rms}
\end{eqnarray}
where the molecular weights are measured in g/mol. The work of
Kinning {\em et al} differs from much of the polymer literature in
that it characterizes the strength of the interaction between the
two polymer species via the {\em interaction energy density}
$\Lambda$ \cite{roe}, rather than the corresponding $\chi$
parameter. The energy of this interaction is given by
\begin{equation}
\Lambda\int\mathrm{d}\mathbf{r}\,\phi_\text{PS}(\mathbf{r})\phi_\text{PB}(\mathbf{r})
\label{Lambda_interaction}
\end{equation}
where the $\phi_i(\mathbf{r})$ terms are the local volume fractions
of the two polymer species at position $\mathbf{r}$. The definition
of the $\chi$ parameter involves the introduction of an arbitrary
reference volume $V_\text{ref}$, such as the average of the repeat
unit volumes of the two polymers \cite{matsen_book}. However, it is
always possible to write terms involving the interaction strength in
such a way that the numerical value of $V_\text{ref}$ need not be
specified. Thus $\Lambda$, which requires no reference volume, is
preferred by some authors. For reference, $\chi$ and $\Lambda$ are
related by \cite{roe}
\begin{equation}
\Lambda = \chi(V_\text{ref}/k_\text{B}T)^{-1} \label{Lambda_chi}
\end{equation}
where $k_\text{B}$ is Boltzmann's constant.

Data on the temperature dependence of the polystyrene-polybutadiene
interaction energy density may be fitted by the empirical formula
\cite{roe_zin}
\begin{equation}
\Lambda(\text{cal}/\text{cm}^3)=A-B(T-150^\circ\text{C})
\label{Lambda_T}
\end{equation}
where $A=0.718\pm 0.051$ and $B=0.0021\pm 0.00045$.

The parameters described here (molecular volumes, end-to-end
distances and interaction energy density) are the required input for
our SCFT calculations. They are all determined from experiments that
do not involve micelle formation -- it is not necessary for any
parameters specific to micellization to be measured in order to
predict the structure of micelles.

To our knowledge, the only approach that has previously been used to
provide detailed, system-specific predictions of micelle structure
in copolymer/homopolymer blends is scaling theory. In the paper of
Kinning {\em et al} discussed above, the experimental data is
compared with a scaling approach developed by Leibler and others
\cite{leibler, whitmore_noolandi, mayes_delacruz, roe}. Like SCFT,
this theory predicts the micelle properties from the system
parameters introduced above. Since scaling theory will be compared
with our SCFT calculations later in this paper, we outline it
briefly now.

Scaling theory assumes that the micelle consists of two distinct and
uniform regions: the core and the corona, which in turn are
completely distinct from the surrounding solution. The core is taken
to contain B-type (solvophobic) monomers mixed with A homopolymer.
Similarly, the corona is assumed to be composed of A-type monomers
(belonging to the copolymer molecules) mixed with A homopolymer. The
surroudings of the micelle are modeled as a homogeneous solution of
homopolymer and copolymer chains. Making these clear divisions
between the different regions of the micelle and the surrounding
polymer solution allows the free energy of a blend containing
$N_\text{m}$ micelles to be written as
\begin{equation}
F_\text{M}=N_\text{m}F+F_\text{mix}-TS_\text{m}
\label{scaling_total}
\end{equation}
where $F$ is the free energy of a single micelle, $F_\text{mix}$ is
the free energy of mixing of the homopolymers and diblocks {\em
outside} the micelles, and $S_\text{m}$ is the translational entropy
of the gas of micelles. $F_\text{mix}$ is calculated from the
standard Flory-Huggins expression \cite{jones_book}. The
translational entropy $S_\text{m}$ is calculated from a lattice
model \cite{leibler, mayes_delacruz} where each lattice cell has the
volume of a single micelle \cite{mayes_delacruz}. The number of
occupied cells is calculated from the fraction of copolymer chains
that aggregate into micelles.

The free energy of a single micelle $F$ can then be split into
separate contributions as follows:
\begin{equation}
F=4\pi R_\text{c}^2\gamma+F_\text{d}+F_\text{mA}+F_\text{mB}
\label{scaling_single}
\end{equation}
Here, $R_\text{c}$ is the radius of the core and $\gamma$ is the
interfacial tension at the core-corona boundary. The use of this
expression for the interfacial energy assumes that the interface has
the same properties as the interface between two incompatible
homopolymers and allows $\gamma$ to be simply expressed in terms of
the monomer length and the $\chi$ parameter \cite{leibler}.

The {\em deformation energy} $F_\text{d}$ arises from the stretching
of the copolymer chains confined to the spherical micelle. Finally,
$F_\text{mA}$ and $F_\text{mB}$ are the free energies of mixing of
the homopolymer with the copolymer blocks in the corona and core
regions respectively. Like the bulk mixing term $F_\text{mix}$ in
Equation \ref{scaling_total}, these are assumed to be given by the
Flory-Huggins theory of polymer blends \cite{jones_book}.

To find the equilibrium state of the system, Equation
\ref{scaling_total} is minimized, keeping the temperature and total
volume fraction of copolymer fixed and allowing all other quantities
(such as the size and number of micelles) to vary. This yields
predictions for the core and corona radii, the critical micelle
concentration and the composition of the various regions. In this
paper, we compare our SCFT results with the scaling predictions of
Kinning {\em et al} \cite{kinning}.

\section{Self-consistent field theory}\label{scft}

Self-consistent field theory (SCFT) \cite{edwards} predicts the
equilibrium structures formed in a melt or blend of polymers. It is
a mean-field theory, neglecting composition fluctuations. The
description of the polymers is coarse-grained: the configuration of
an individual polymer molecule is modeled as a random walk in space
$\mathbf{r}_\alpha(s)$, where $s$ is a curve parameter specifying
the position along the polymer backbone. An ensemble of many such
polymers is considered. The interactions between polymers are
included in two ways: by assuming that the blend is incompressible
and introducing a contact potential between molecules of different
species. The strength of this potential is specified by the
interaction energy density $\Lambda$.

The first step in finding an approximation method (SCFT) for the
above system is to view each molecule as being acted on by a field
produced by all other molecules in the system \cite{matsen_book}.
This transforms the $N$-body problem of modeling an ensemble of $N$
polymers into $N$ $1$-body problems. Since we wish to compute the
partition sum over all possible system configurations, all molecules
of a given species may be treated as equivalent. Therefore, we need
only introduce one field $W_i(\mathbf{r})$ for each species $i$ and
have only to solve $i$ $1$-body problems. Note that this involves no
approximation -- the complexity of the system is hidden in the field
terms $W_i(\mathbf{r})$. The utility of the SCFT approach lies in
the fact that approximations may be found more easily for the fields
than for the original formulation of the problem.

For definiteness, we now focus on our diblock copolymer/homopolymer
blend, and introduce fields $W_\text{PS}$, $W_\text{PB}$ and
$W_\text{hPS}$ acting on the polystyrene blocks, polybutadiene
blocks and homopolystyrene respectively. This converts the partition
sum into an integral over field configurations, with the original
Hamiltonian replaced by an effective Hamiltonian $H$. Following the
procedure described in the recent review by Matsen
\cite{matsen_book}, we find that $H$ is given by
\begin{eqnarray}
\lefteqn{\frac{H}{k_\text{B}T} =
\frac{\Lambda}{k_\text{B}T}\int\mathrm{d}\mathbf{r}\,[\Phi_\text{PS}(\mathbf{r})+\Phi_\text{hPS}(\mathbf{r})][1-\Phi_\text{PS}(\mathbf{r})-\Phi_\text{hPS}(\mathbf{r})]}\nonumber\\
& &
-\frac{1}{v_\text{PS}+v_\text{PB}}\int\mathrm{d}\mathbf{r}\,\{W_\text{PS}(\mathbf{r})\Phi_\text{PS}(\mathbf{r})+W_\text{PB}(\mathbf{r})[1-\Phi_\text{PS}(\mathbf{r})-\Phi_\text{hPS}(\mathbf{r})]\}\nonumber\\
& &
-\frac{1}{v_\text{hPS}}\int\mathrm{d}\mathbf{r}\,W_\text{hPS}(\mathbf{r})\Phi_\text{hPS}(\mathbf{r})-\frac{\overline{\phi}_\text{hPS}V}{v_\text{hPS}}\ln
Q_\text{hPS}-\frac{(\overline{\phi}_\text{PS}+\overline{\phi}_\text{PB})V}{v_\text{PS}+v_\text{PB}}\ln
Q_\text{PS,PB} \label{hamiltonian}
\end{eqnarray}
where the $\Phi_i(\mathbf{r})$ are the local volume fractions of the
various polymer species $i$ ($i=\text{PS}$, PB or hPS) and $V$ is
the total volume of the system. The terms $\overline{\phi}_i$ are
the mean volume fractions of species $i$. The first term (with
prefactor $\Lambda/k_\text{B}T$) describes the interaction between
the different polymer species. In this term, and throughout, the
polybutadiene block volume fraction $\Phi_\text{PB}(\mathbf{r})$ is
replaced by
$1-\Phi_\text{PS}(\mathbf{r})-\Phi_\text{hPS}(\mathbf{r})$. This
follows from the incompressibility of the blend: the local volume
fractions must add to $1$ at every point. The next three terms all
involve the $W_i(\mathbf{r})$, and arise from the unit operators
that are inserted into the partition function to convert the
partition sum into an integral over fields. In the penultimate term,
$Q_\text{hPS}$ is the partition function of a single homopolymer
chain in the field $W_\text{hPS}(\mathrm{r})$. Similarly,
$Q_\text{PS,PB}$ is the partition function of a single copolymer
chain in the fields $W_\text{PS}(\mathrm{r})$ and
$W_\text{PB}(\mathrm{r})$. These are given by (again following the
review referenced above \cite{matsen_book})
\begin{equation}
Q_\text{hPS}[W_\text{hPS}]=\int\mathrm{d}\mathbf{r}\,q_\text{hPS}(\mathbf{r},s)q_\text{hPS}^\dagger(\mathbf{r},s)
\label{single_chain_partition}
\end{equation}
where the $q$ and $q^\dagger$ terms are single chain propagators
\cite{matsen_book}. The partition function $Q_\text{PS,PB}$ of a
single copolymer chain is determined similarly. Recall that the
polymer molecules are modeled as random walks subject to an external
field that incorporates their interactions with the rest of the
melt. This is reflected in the fact that the propagators satisfy
modified diffusion equations. In the case of the homopolymer, we
have
\begin{equation}
\frac{\partial}{\partial
s}q_\text{hPS}(\mathbf{r},s)=\left[\frac{1}{6}\langle
R^2_\text{hPS}\rangle\nabla^2-W_\text{hPS}(\mathbf{r})\right]q_\text{hPS}(\mathbf{r},s)
\label{diffusion}
\end{equation}
with initial condition $q_\text{hPS}(\mathbf{r},0)=1$. The copolymer
propagators are computed in a similar way, with the copolymer
architecture entering into the theory through the fact that the
corresponding diffusion equation for the copolymer is solved with
the field $W_i(\mathbf{r})$ and the prefactor of the $\nabla^2q$
term appropriate to each of the two sections of the copolymer
\cite{fredrickson_book}.

All steps until now have been exact, and we now introduce the main
approximation of SCFT. This consists of extremizing $H$ with
respect to all fields $W_i(\mathbf{r})$ and all densities
$\Phi_i(\mathbf{r})$, yielding a saddle-point approximation to the
system partition function $Z$. If the polymers are long and
fluctuations are weak, this method successfully isolates the
dominant contribution to $Z$, and the predictions of SCFT are in
good agreement with experiment \cite{bates_fredrickson_rev}.

Extremizing $F$ produces a set of simultaneous equations linking
the values of the fields and densities at the minimum. Denoting
these values by lower-case letters $\phi_i(\mathbf{r})$ and
$w_i(\mathbf{r})$, we find
\begin{eqnarray}
1&=&\phi_\text{PS}(\mathbf{r})+\phi_\text{PB}(\mathbf{r})+\phi_\text{hPS}(\mathbf{r})\nonumber\\
\frac{1}{v_\text{PS}+v_\text{PB}}\left[w_\text{PS}(\mathbf{r})-w_\text{PB}(\mathbf{r})\right]&=&\frac{2\Lambda}{k_\text{B}T}\left[\overline{\phi}_\text{PS}+\overline{\phi}_\text{hPS}-\phi_\text{PS}(\mathbf{r})-\phi_\text{hPS}(\mathbf{r})\right]\nonumber\\
w_\text{hPS}&=&\frac{v_\text{hPS}}{v_\text{PS}+v_\text{PB}}w_\text{PS}(\mathbf{r})
\label{SCFT_equations}
\end{eqnarray}
where $\overline{\phi}_i$ is the (mean) volume fraction of species
$i$. The first of these equations arises directly from the
incompressibility of the melt. The homopolymer density is evaluated
from the propagators (see Equation \ref{diffusion}) according to
\cite{matsen_book}
\begin{equation}
\phi_\text{hPS}(\mathbf{r})=\frac{V\overline{\phi}_\text{hPS}}{Q_\text{hPS}[w_\text{hPS}]}\int^1_0\mathrm{d}s\,
q_\text{hPS}(\mathbf{r},s)q_\text{hPS}^\dagger(\mathbf{r},s)
\label{density}
\end{equation}
The copolymer densities are calculated similarly, with the
integration limits set to give the correct proportions of each
species.

In order to calculate the SCFT density profiles for a given set of
mean volume fractions, the set of simultaneous equations
\ref{SCFT_equations} must be solved with the densities calculated as
in Equation \ref{density}. This is achieved using a standard
relaxation procedure \cite{matsen2004}. To begin, we make a guess
for the form of the fields $w_i(\mathbf{r})$ and solve the diffusion
equations to calculate the propagators and hence the densities
corresponding to these fields (see Equations \ref{diffusion} and
\ref{density}). New values for the fields are now calculated using
the new $\phi_i(\mathbf{r})$. However, if we then recalculate the
densities with these new $w_i$, we find that the relaxation
algorithm becomes unstable. To avoid this problem, we (heavily) damp
the iteration, replacing the $w_i(\mathbf{r})$ with a mixture of the
old and new values ($0.99w_i^\text{old}+0.01w_i^\text{new})$) and
then calculate the $\phi_i$. The procedure is repeated until the
difference between the left and right hand sides of equations
\ref{SCFT_equations} is less than $10^{-5}$. For several systems, we
have confirmed that the iteration converges to the same solution for
different initial guesses for the $w_i$.

Since we focus on isolated spherical micelles, the diffusion
equations themselves are solved in a spherically-symmetric geometry.
Reflecting boundary conditions are imposed at the origin and at the
boundary of the system. A finite difference method \cite{num_rec} is
used to generate approximate numerical solutions to the diffusion
equations. The calculations are performed in real space, in contrast
to much of the SCFT literature \cite{cavallo}. A step size of
$\Delta r= 4\,\text{\AA}$ is used. It has been checked that using a
finer grid does not significantly change the $\phi_i(\mathbf{r})$.

In the above discussion, we considered a simple system of fixed
volume and fixed copolymer volume fraction containing one micelle.
To find the micelle with the lowest free energy, we must consider
how a system of many micelles minimizes its free energy. Consider a
macroscopic copolymer/homopolymer blend whose copolymer volume
fraction $\bar{\phi}_{PB} + \bar{\phi}_{PS}$, total volume $V_T$ and
temperature $T$ are all fixed. The equilibrium state of this system
can be found by minimizing the total free energy $F$. If the
copolymer concentration is above a certain value (the \emph{critical
micelle concentration}), copolymer chains can either exist as
monomers or in micelles. The number density of micelles is thus an
internal degree of freedom and the macroscopic system varies this
quantity (subject to the constraint of fixed copolymer volume
fraction) in order to minimize the free energy.

Explicit calculations on this many-micelle macrosopic system are
extremely time-consuming even using SCFT. However, we can reduce the
problem to one involving only a single micelle if we neglect
inter-micellar interactions and the translational entropy of the
micelles. The former is applicable if the micellar solution is
sufficiently dilute while the latter introduces a (small) correction
term to the free energy which will be included by hand later in this
section. In this case, we can reduce the many-micelle system to a
one-micelle system of volume $V$ and copolymer volume fraction
$\bar{\phi}_{PB} + \bar{\phi}_{PS}$, where $V$ corresponds to the
volume per micelle. We can then effectively vary the number density
of micelles by varying $V$. Specifically, if the free energy of this
subsystem is $A$, the total free energy of the macroscopic system is
given by $F=\frac{V_T}{V} A$. Since $V_T$ is fixed, we can find the
equilibrium state of the whole system by minimizing the free energy
density $a = A/V$ with respect to $V$. In the Appendix, we show
using a simple two state model that this procedure automatically
yields the optimum micelle, that is the micelle with the lowest free
energy per chain\cite{safran_book}.

To our knowledge, this method of varying the size of the calculation
box containing a single micelle in order to obtain information on a
system of many micelles has not been used before: in earlier work,
the box size is fixed \cite{vanlent}.
Although having to minimize $a$ with respect to $V$ for each system
parameter adds to the numerical burden of the calculation, the
advantage of this method is that, by minimizing the free energy
density, we avoid the awkward problem of trying to define the free
energy per chain in the micelle, which is the basic quantity in
simple theories of micellization \cite{safran_book}. Taking this
latter approach would involve making ad hoc definitions concerning
the boundary of the micelle in what is essentially a continuum
calculation.

Another advantage of our approach is that it yields a well-defined
value for the volume occupied by a micelle. This allows us to take
into account the translational entropy of spherical
micelles\cite{mayes_delacruz}. An estimate of the translational entropy per
micelle can be obtained from a simple lattice model where the system
is divided into cells of the volume of a single micelle. We adapt
the results of Mayes and de la Cruz \cite{mayes_delacruz} and Leibler \cite{leibler} and find that
the translational entropy per micelle is
\begin{equation}\label{trans entropy}
    S_{trans} = -k_B \left[ \ln
    \left(\frac{V_m}{V}\right)+\left(\frac{V-V_m}{V_m}\right)\ln \left(\frac{V-V_m}{V}\right)
    \right]
\end{equation}
where $V_m$ is the volume of the micelle and $V$ is the volume of
the subsystem containing the micelle. Note that the lattice model
leading to Equation \ref{trans entropy} implicitly assumes that
micelles are impenetrable. Equation \ref{trans entropy} thus also
partially corrects for inter-micellar interactions which were
neglected in the preceding discussion. To estimate the micelle
volume, we use the working definitions of the core radius $R_c$ and
the corona thickness $L_c$ explained in the next section; $V_m$ can
then be calculated directly from $R_c+L_c$.

We now have all the necessary techniques to calculate the optimum
micelle for each system parameter. To begin, we perform an SCFT
calculation at fixed subsystem volume, giving the density profile of
a micelle and the free energy density of the subsystem. We then
adjust the subsystem volume. This is achieved by changing the number
of points on the grid on which we solve the diffusion equations
whilst keeping the grid step size constant. This is repeated until
we have located the minimum of the free energy density for the
geometry under consideration.

This procedure yields the density profile of the micelle with the
minimum free energy per chain $F_\text{chain}$. As shown in the
Appendix, it also results in the chemical potential of bulk
copolymers outside the micelle being equal to $\min\{F_{chain}\}$.
Our approach therefore also allows us to estimate the critical
micelle concentration (more accurately the critical micelle volume
fraction) which we define via the equation \cite{safran_book}
\begin{equation} \label{safran_cmc}
\phi_c-P_1=P_1 \Rightarrow \phi_c=2 P_1
\end{equation}
where $P_1$ is the bulk copolymer concentration that coexists with
the micelle at total copolymer fraction $\phi$. Note that, in the
original definition \cite{safran_book}, $P_1$ is the total volume
fraction of copolymers that exist as isolated molecules, rather than
the bulk copolymer concentration that coexists with the micelle.
However, provided the micelles are well-separated, these two
quantities are very close and we shall treat them as equivalent.
Strictly speaking, in the above definition for $\phi_c$, we should
also use the bulk copolymer concentration \emph{at CMC} rather than
at $\phi$. This is because the CMC is defined as the value of $\phi$
where the volume fraction of micelles equals the volume fraction of
bulk copolymer. However in practice using the bulk copolymer
fraction at $\phi$ rather than at CMC yields a very good
approximation for $\phi_c$ since the bulk copolymer volume fraction
remains pinned close to the CMC value for $\phi > \phi_c$. This is
most readily illustrated using a simple bimodal model for micelle
formation where we assume that copolymers can exist either as single
molecules or as micelles with the optimum aggregation number $M$
\cite{safran_book}. Denoting the micelle and single molecule
fraction at $\phi$ as $P_M$ and $P_1$ respectively and the micelle
and single molecule fraction at $\phi_c$ as $P_{Mc}$ and $P_{1c}$
respectively, it is easy to show that for $\phi >> \phi_c$,
$P_1=P_{1c} (\phi/\phi_c)^{1/M}$. Since $M$ is typically large, the
volume fraction of copolymers outside the micelle remains pinned
close to its CMC value. For comparison with experimental data,
$\phi_c$ is converted to weight fraction using specific volumes
(Equations \ref{V_PB} and \ref{V_PS}).

In the results presented in next section, the core and corona
thicknesses of the spherical micelle are calculated at 10\% weight
fraction of copolymer, i.e., the weight fraction used in the
original experiments \cite {kinning}. However the CMC values are
calculated at 2.5\% weight fraction of copolymer. For the copolymers
with the lighter core blocks (SB 10/10, SB 40/10), the choice of
weight fraction makes relatively little difference to the calculated
CMC, as we would expect from the argument in the previous paragraph.
However for the copolymers with the heavier core block (SB 20/20),
we need to use the dilute (2.5\% weight fraction) system in order
for the copolymer concentration to have reached a constant value at
the edge of the system. Decreasing the copolymer weight fraction further
has only a very small effect on our prediction for the cmc.

\section{Results and discussion}\label{results}
\begin{figure}
\includegraphics[width=\linewidth]{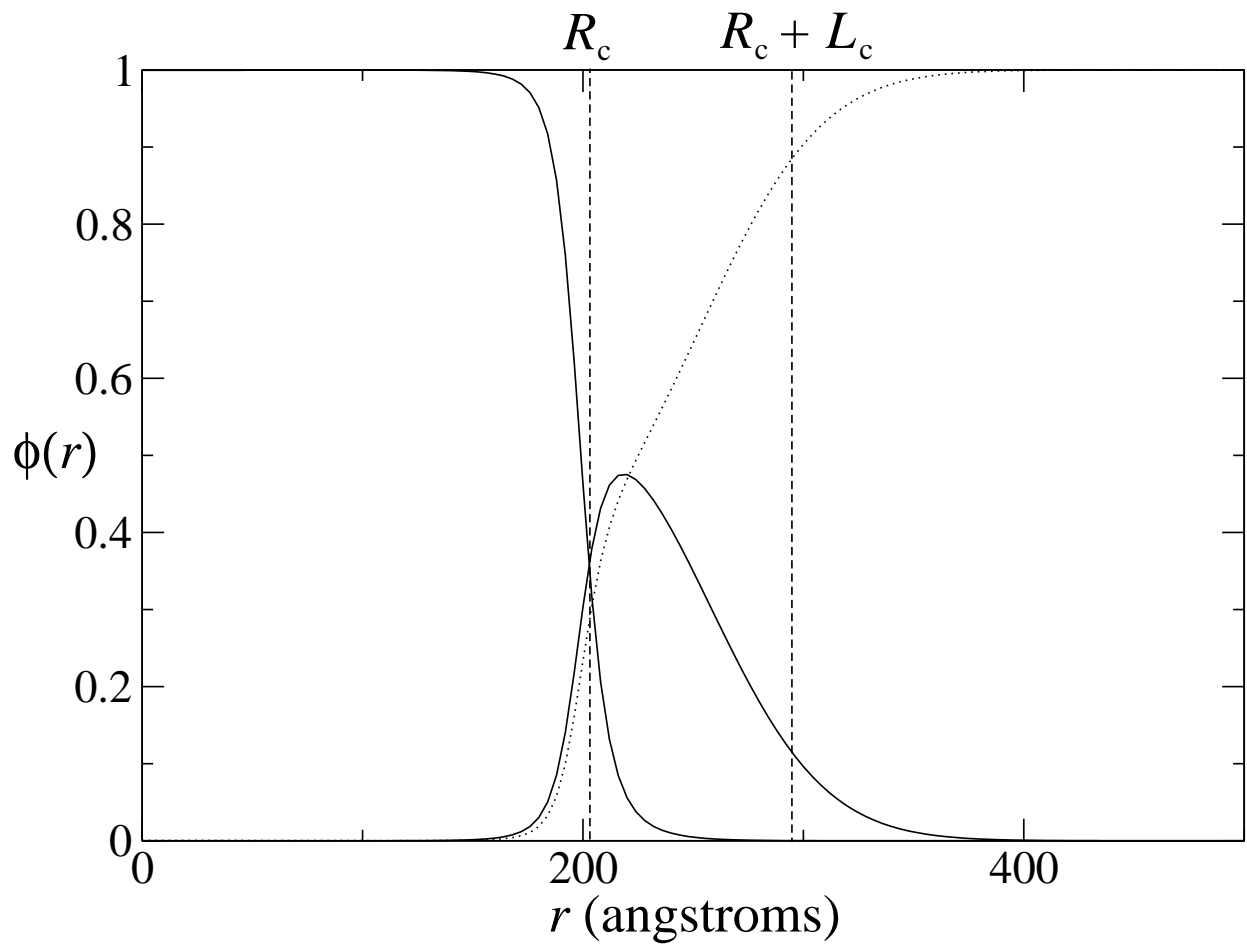}
\caption{\label{cross_section_fig} Cross-section showing the volume
fraction profiles of the various species in a spherical micelle at
$115^\circ\text{C}$ as a function of the distance $r$ from the
centre of the micelle. Full line: volume fraction profile for PB
blocks; dashed line: PS; dotted line: homopolystyrene. The core
radius $R_\text{c}$ and the corona thickness $L_\text{c}$ are also
marked. }
\end{figure}

We begin by comparing our predictions for the core radii and corona
thicknesses with experiment and scaling theory for a variety of
poly(styrene-butadiene)/homopolystyrene blends. In contrast to
scaling theory, the borders between the core, corona and bulk
regions are not perfectly sharp in SCFT, and we require working
definitions of the core radius $R_\text{c}$ and corona thickness
$L_\text{c}$. We define the core radius as that at which the volume
fractions of the solvophobic and solvophilic blocks are equal:
$\phi_\text{PB}(\mathbf{r})=\phi_\text{PS}(\mathbf{r})$. Other
definitions, such as
$\phi_\text{PB}(\mathbf{r})=\phi_\text{PS}(\mathbf{r})+\phi_\text{hPS}(\mathbf{r})$
\cite{katsov1} are equally valid. However, since the boundaries are
quite sharp, the difference between the two definitions for a given
system is quite small (see \ref{cross_section_fig}).

\begin{figure}
\includegraphics[width=\linewidth]{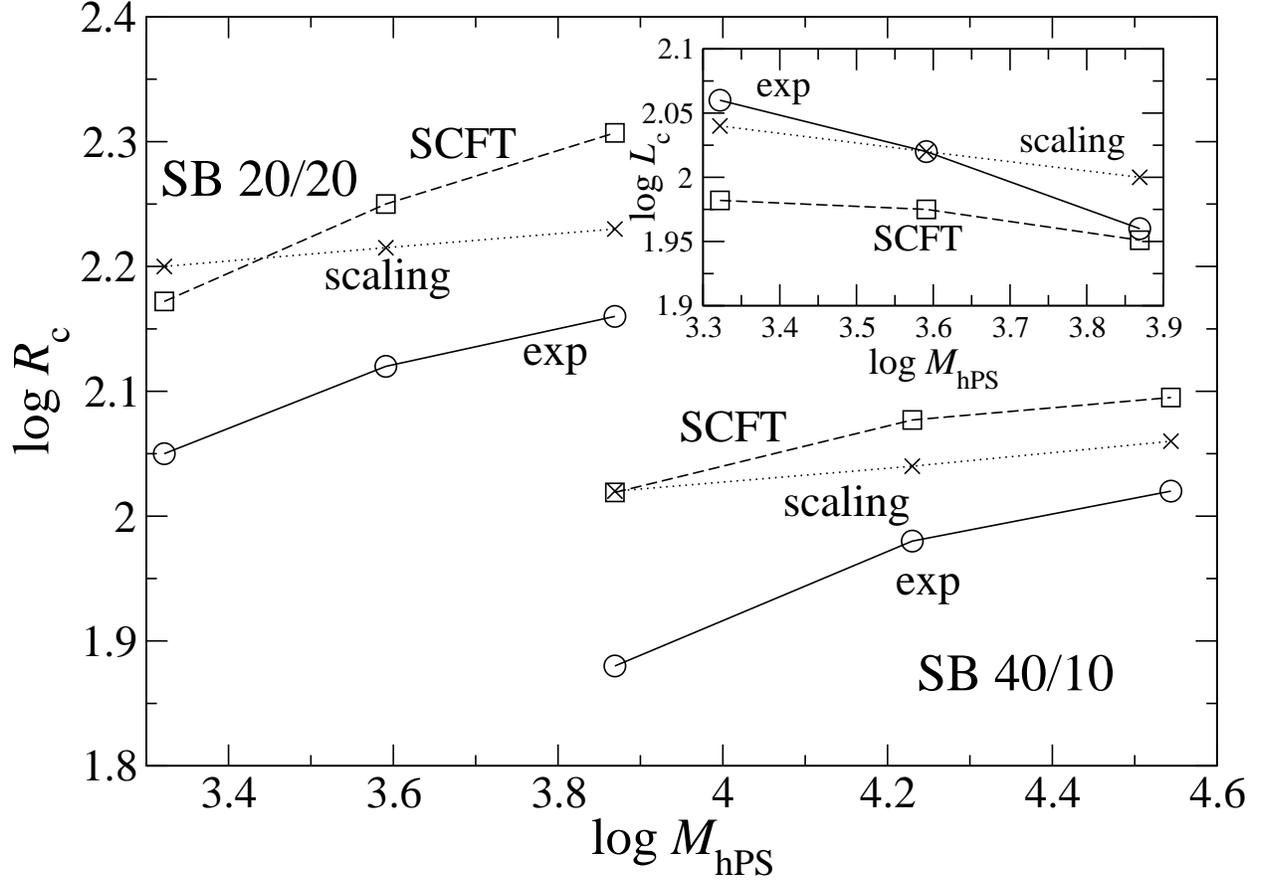}
\caption{\label{coreMhPS_fig} The main panel shows a comparison of
SCFT predictions with experiment and scaling at $115^\circ\text{C}$
and copolymer weight fraction $10\,\%$
for core radii $R_\text{c}$, as a function of homopolystyrene
molecular weight. The data in the top left are for copolymers SB
20/20; those in the bottom right are for SB 40/10. The inset shows
corona thickness $L_\text{c}$ as a function of homopolystyrene
molecular weight for SB 20/20. In both the main panel and the inset,
solid lines with circles show experimental results, dashed lines
with squares show SCFT predictions and dotted lines with crosses
show scaling. }
\end{figure}

In contrast, the corona is very diffuse (see
\ref{cross_section_fig}), and it is less clear how to define its
outer boundary. To proceed, we calculate the radius of gyration of
the corona from \cite{burchard}
\begin{equation}
R_\text{g}^2=\frac{\int r^2
(\phi_\text{PS}(r)-\phi_\text{PS}^\text{b})4\pi
r^2\mathrm{d}r}{\int(\phi_\text{PS}(r)-\phi_\text{PS}^\text{b})4\pi
r^2\mathrm{d}r} \label{R_g_def}
\end{equation}
where $\phi_\text{PS}^\text{b}$ is the bulk polystyrene
concentration which must be removed to isolate the corona. We now
calculate the thickness $L_\text{c}$ of the hollow sphere with inner
radius $R_\text{c}$ (the core radius as defined above) which has the
same radius of gyration as the corona. This is taken as an estimate
of the corona thickness. $L_\text{c}$ is related to $R_\text{c}$ and
$R_\text{g}$ by \cite{burchard}
\begin{equation}
R_\text{g}^2=\frac{3}{5}\frac{(R_\text{c}+L_\text{c})^5-R_\text{c}^5}{(R_\text{c}^3+L_\text{c}^3)-R_\text{c}^3}
\label{R_L_R}
\end{equation}
and can be determined numerically.

We now examine the effect of homopolymer molecular weight on core
and corona size for the samples SB 20/20 and SB 40/10
\cite{kinning}. The experimental data (alongside scaling theory and
our SCFT predictions) are shown in \ref{coreMhPS_fig} as a plot of
the logarithm of the core radius against the logarithm of the
homopolymer molecular weight. As the homopolystyrene molecular
weight is decreased at a fixed polystyrene block molecular weight,
we see that the corona thickness increases and the core radius
decreases. This is because decreasing homopolystyrene length leads
the entropy of mixing between the homopolystyrene and the
polystyrene block to increase, increasing the tendency of the
homopolystyrene volume fraction profile to be uniform. This can be
achieved by making the corona region larger and the core smaller
\cite{kinning}. These data also show the effect of core block
molecular weight: if a copolymer with a larger PB block (SB 20/20
rather than SB 40/10) is used, the core radius is naturally larger.

The shape of these experimentally-observed trends are reproduced
much more accurately by SCFT (see \ref{coreMhPS_fig}) compared to
scaling theory, where the predicted variation of $R_c$ with
$M_{hPS}$ is too shallow. On the other hand for both SB 20/20 and SB
40/10, SCFT overestimates $R_c$ by about 20-30\%; this discrepancy
is comparable to that found in scaling theory. However, we emphasize
that our calculations contain no adjustable parameters: all the
required input concerning polymer properties (such as the
interaction density $\Lambda$) has been determined from experiments
that do not involve micelle formation. Given this fact, we believe
that the agreement between SCFT and experiment for $R_c$ is good.

With our corona definition, SCFT reproduces the experimentally
observed fall in corona thickness as the homopolymer molecular
weight is increased for the sample SB 20/20 (see inset to
\ref{coreMhPS_fig}). However both SCFT and scaling theory predict a
variation of $L_c$ with $M_{hPS}$ that is too shallow. In terms of
the magnitude of $L_c$, scaling theory is marginally more accurate
compared to SCFT, though the difference between either SCFT or
scaling theory with experiment is small. We also note that the
corona is very diffuse and its radius is measured in experiments by
an indirect procedure (it is inferred from the inter-particle
interactions) \cite{kinning_thomas_fetters}. The measurements of the
corona thickness do not therefore constitute as stringent a test of
the theory as those of the core radius.

In terms of the variation of $R_c$ or $L_c$ with $M_{hPS}$,
experimental data are available for heavier copolymers (SB 80/80)
\cite{kinning}. Unfortunately we have not been able to test SCFT for
this system because $\chi N$ is too high and the core density
profile too sharp so that our extremization procedure becomes
unreliable due to numerical errors.

\begin{figure}
\includegraphics[width=\linewidth]{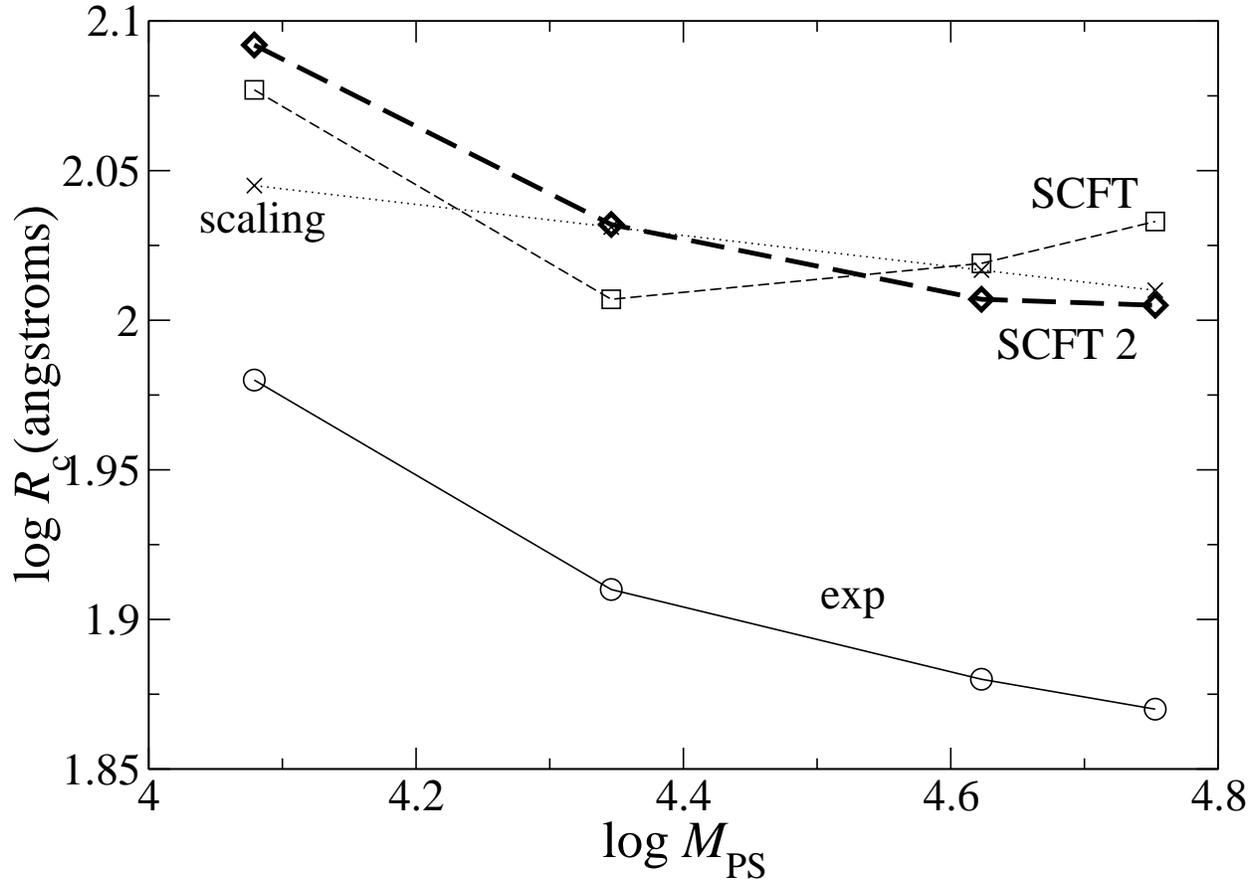}
\caption{\label{coreMPS_fig} Comparison of SCFT predictions (dashed
lines with squares) with experiment (solid lines with circles) and
scaling (dotted lines with crosses) at $115^\circ\text{C}$ and $10\,\%$
copolymer weight fraction for core
radii $R_\text{c}$, as a function of polystyrene block molecular
weight. The molecular weight of the polybutadiene (core) block is
held approximately constant. The line labeled SCFT1 uses the core block
molecular weights used in the experiment; the bolder line labeled SCFT2 
fixes the core block molecular weight at 10 kg/mol. See the text for a
discussion.}
\end{figure}

Next, we consider the dependence of the core radius on the
polystyrene block molecular weight (rather than the homopolystyrene
weight considered above). As before, the SCFT predictions, scaling
theory and experimental data are shown in a log-log plot
(\ref{coreMPS_fig}). The PB block molar weight is fixed at
approximately 10 kg/mol, whilst the PS block molar weight is
increased from around 10 kg/mol to 60 kg/mol. The homopolymer is
always 7400PS. From the experimental data, the PB core is seen to
shrink as the PS block molecular weight is increased. This behavior
can be understood (see the explanation given above) by considering
the effect of the polystyrene molecular block weight (this time at
fixed homopolystyrene weight) on the entropy of mixing outside the
core \cite{kinning}.

For the SCFT curve labeled "SCFT 1", the PB block molar weight for
each copolymer sample is fixed at the value quoted in the original
paper\cite{kinning}, i.e., the copolymer samples considered in
\ref{coreMPS_fig} do not all have the same PB block molar weight,
but instead have a range of molar weights in the region of 10
kg/mol. In this case, SCFT does not predict the shape of $R_c$ vs
$M_{PS}$ very well. In particular the predicted $R_c$ is no longer a
monotonically decreasing function of $M_{PS}$. The scatter in the
SCFT points comes from the fact that within SCFT, $R_c$ is very
sensitive to the exact molecular weight of the core block so any
scatter in the input PB block weight is amplified in the final $R_c$
results. In particular, the sample SB 23/10 (the second point in
\ref{coreMPS_fig}) has the rather lower value of 9 kg/mol which
leads to an apparent minima in the $R_c$ values.

If on the other hand we fix the core block for all the samples to be
10 kg/mol, we obtain the curve labeled "SCFT 2" which now predicts
the shape of $R_c$ vs $M_{PS}$ much more accurately. In fact, fixing
the PB block weight of all the samples to be 10 kg/mol lies well
within the experimental accuracy of the measured molecular weights
since the polydispersity for all the copolymer samples can be as high
as $M_w/M_n=1.05$ \cite{kinning_thomas_fetters} which translates to an uncertainty in
the molecular weight of $\langle\Delta M^2 \rangle^{1/2}/M_n
\approx 20\%$ (recall \cite{strobl} that $\langle\Delta M^2
\rangle^{1/2}/M_n=M_w/M_n-1$). We note that the
curve SCFT2 predicts the shape of $R_c$ vs $M_{PS}$ much more
accurately than scaling theory, where the predicted variation of
$R_c$ with $M_{PS}$ is too shallow. However as before, SCFT
overestimates $R_c$ by about 30\%, similar to scaling theory.

\begin{figure}
\includegraphics[width=\linewidth]{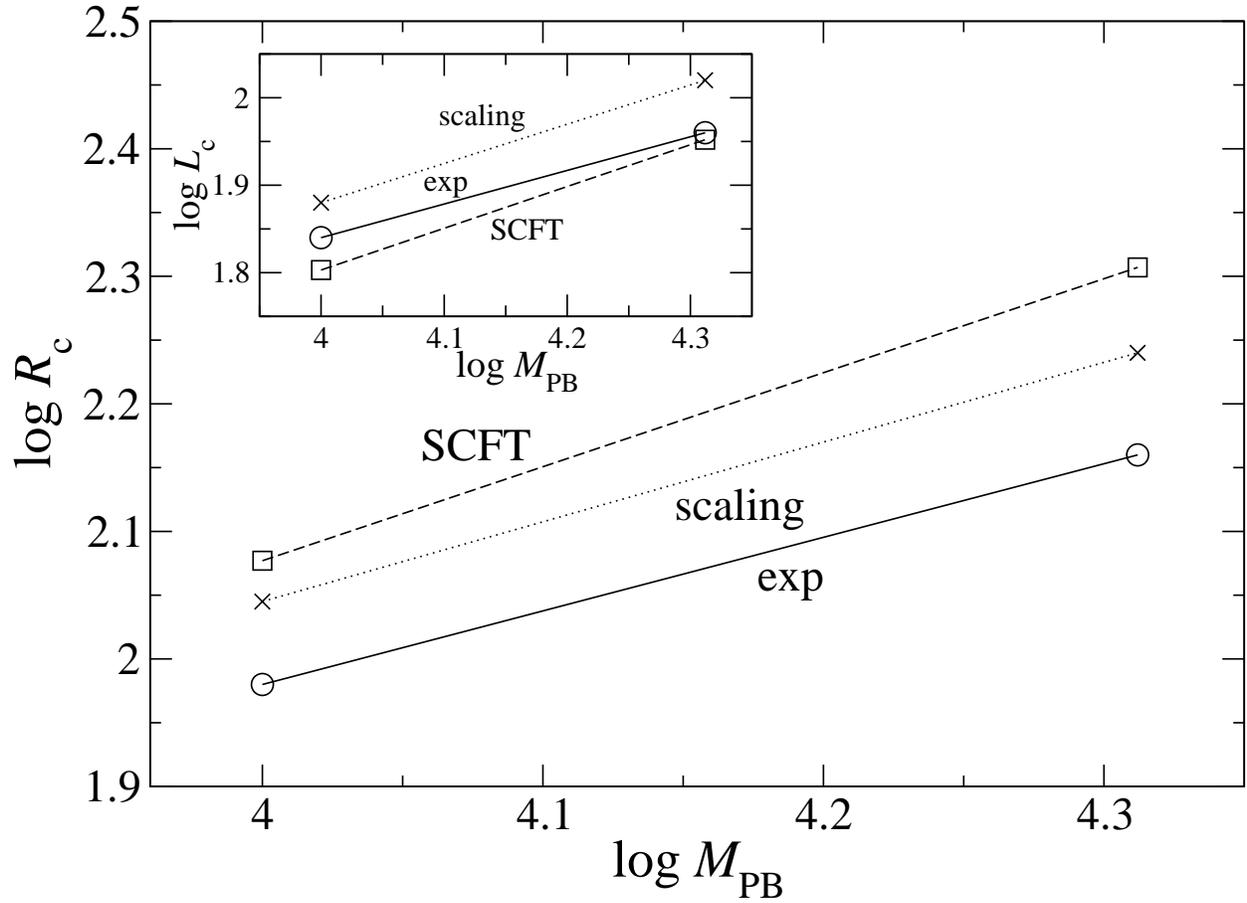}
\caption{\label{coreMPB_fig} Comparison of SCFT predictions (dashed
lines with squares) with experiment (solid lines with circles) and
scaling (dotted lines with crosses) at $115^\circ\text{C}$, as a
function of polybutadiene block molecular weight for symmetric
copolymers. Main panel shows core radii $R_\text{c}$, inset shows
corona thickness $L_\text{c}$. The homopolystyrene molecular weight
is fixed at 7400 g/mol. }
\end{figure}

We now move to the dependence of the core radius and corona
thickness on the overall copolymer molecular weight for {\em
symmetric} copolymers. We consider the samples SB 10/10 and SB 20/20
in 7400PS homopolymer, and, as before, show experimental, scaling
\cite{kinning} and SCFT results in a log-log plot
(\ref{coreMPB_fig}). As would be expected, the core radius and
corona thickness both increase as the molecular weight of the
symmetric diblocks is increased. For the available points, both SCFT
and scaling theory predict the slope of $R_c$ vs $M_{PB}$ well.
However both theories once again overestimate the magnitude of
$R_c$, though scaling theory is marginally more accurate than SCFT
in this case. The situation is reversed for the corona thickness
(inset of \ref{coreMPB_fig}) where SCFT is marginally more accurate
than scaling theory. Unfortunately we are unable to confirm whether
these trends continue to higher copolymer weights because although
experimental data are available for heavier copolymers
\cite{kinning}, we encounter numerical difficulties in modeling
these because of the high $\chi N$ values involved.

\begin{figure}
\includegraphics[width=\linewidth]{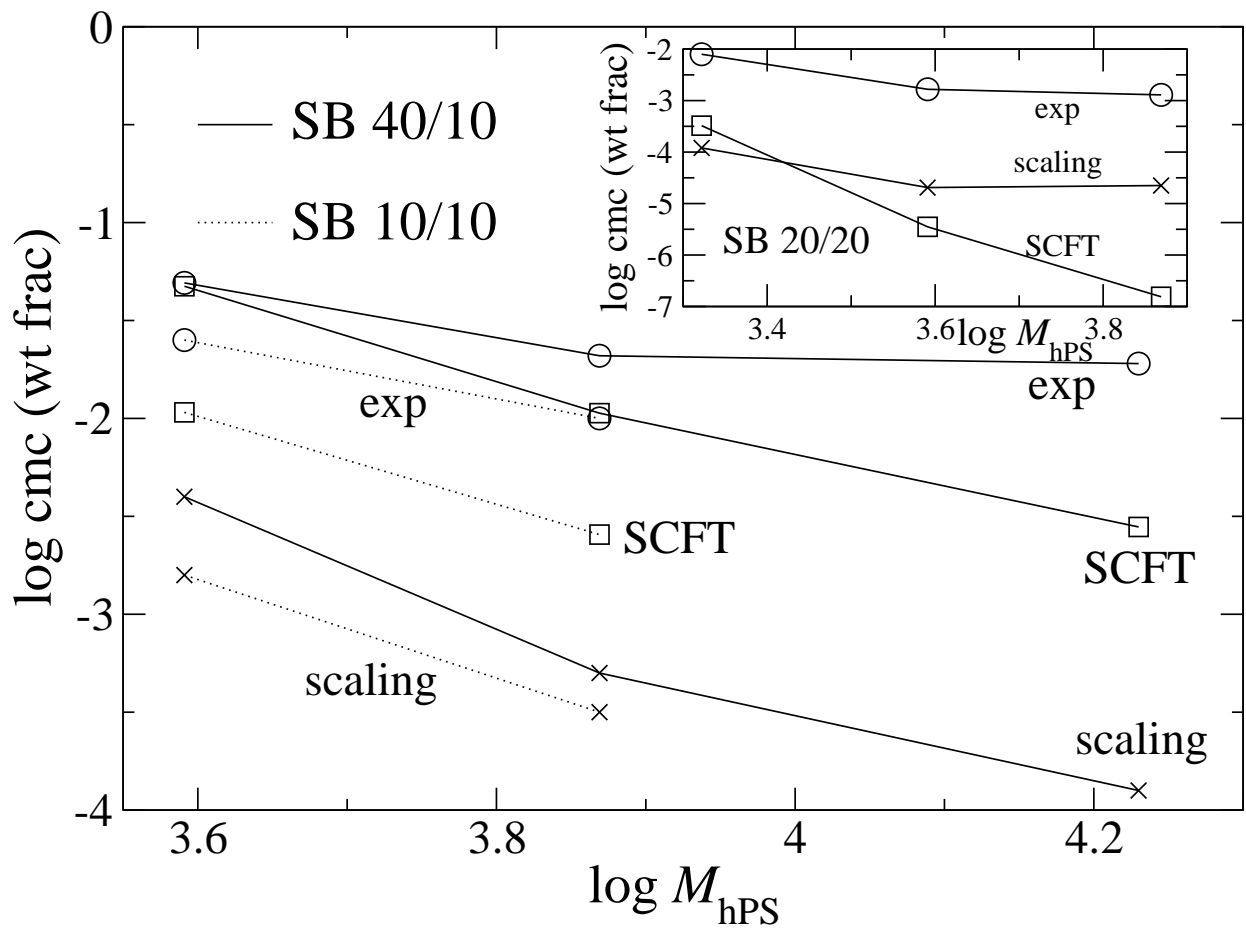}
\caption{\label{cmc_fig} Comparison of SCFT predictions (squares)
with experiment (circles) and scaling (crosses) for the critical
micelle concentration at $115^\circ\text{C}$, as a function of
homopolystyrene molecular weight. Main panel: solid lines show data
for copolymers SB 40/10; dotted lines show results for copolymers SB
10/10. Inset: data for copolymers SB 20/20. }
\end{figure}

Our final comparison of SCFT with experiment and scaling theory
focuses on the data of Kinning {\em et al} concerning the cmc. These
show that the cmc decreases with increasing copolymer molecular
weight (at fixed copolymer composition) and with increasing
homopolystyrene molecular weight. It also increases with increasing
polystyrene block length (see \ref{cmc_fig}). These broad trends are
successfully reproduced by scaling theory \cite{leibler} (see
\ref{cmc_fig}). However, the scaling theory cmcs are 1--2 orders of
magnitude smaller than those measured \cite{kinning}.

We find (see \ref{cmc_fig}) that SCFT also successfully reproduces
the general trends outlined above. Furthermore, for copolymers with
short polybutadiene blocks (SB 10/10 and SB 40/10), SCFT predictions
improve on those of scaling theory by around an order of magnitude,
although the cmc is still underestimated. However, for the copolymer
sample with the longest solvophobic block (SB 20/20), the SCFT
values for the cmc are very low and are in fact inferior to those
predicted by scaling theory (see inset to \ref{cmc_fig}).
Interestingly, similar results were found by Wijmans and Linse
\cite{wijmans_linse}. They compared SCFT with Monte Carlo
simulations for a system of AB diblock copolymers in A {\em monomer}
solvent, and the cmc predicted by SCFT was found to be much lower
than that measured in the simulations (see also the similar
comparison in the simulation and SCFT work of Cavallo {\em et al}
\cite{cavallo}). Wijmans and Linse argued that this was due to the
effect of the mean-field approximation on the bulk chemical
potential. The argument is as follows \cite{wijmans_linse}. In
mean-field theory, the effect of compositional fluctuations where an
individual copolymer interacts with itself are neglected, so that a
copolymer in the bulk region interacts almost exclusively with the
solvent. The solvophobic B section of the copolymer therefore
interacts largely with the A solvent, and the number of repulsive
contacts is very high. However, in Monte Carlo simulations or
experiment, the B block minimizes contact with the solvent by
collapsing into a globule, so that the number of repulsive
interactions is smaller than that predicted by mean-field theory.
This means that the bulk phase is more energetically unfavorable in
SCFT, and micelle formation is hence more favorable in comparison.
The cmc is therefore underestimated by SCFT. If the above argument
is correct, we would expect the disagreement to worsen sharply when
we consider polymers with longer solvophobic blocks. This is indeed
seen for SB 20/20.

Interestingly, the agreement between scaling theory and experiment
for SB 20/20 is not as bad, despite the fact that a mean field
Flory-Huggins model is employed for the bulk free energy
\cite{leibler} (see the inset to \ref{cmc_fig}). We speculate that
the better agreement may be due to a fortuitous cancellation of
errors due to the approximate nature of the free energy used in
scaling theory for both the micelle and the bulk. In contrast, we
expect the SCFT free energy to be very accurate for the micelle (as
evidenced by the excellent agreement of core dimensions) but
significantly poorer for the bulk for the reasons explained in the
previous paragraph.

\begin{figure}
\includegraphics[width=\linewidth]{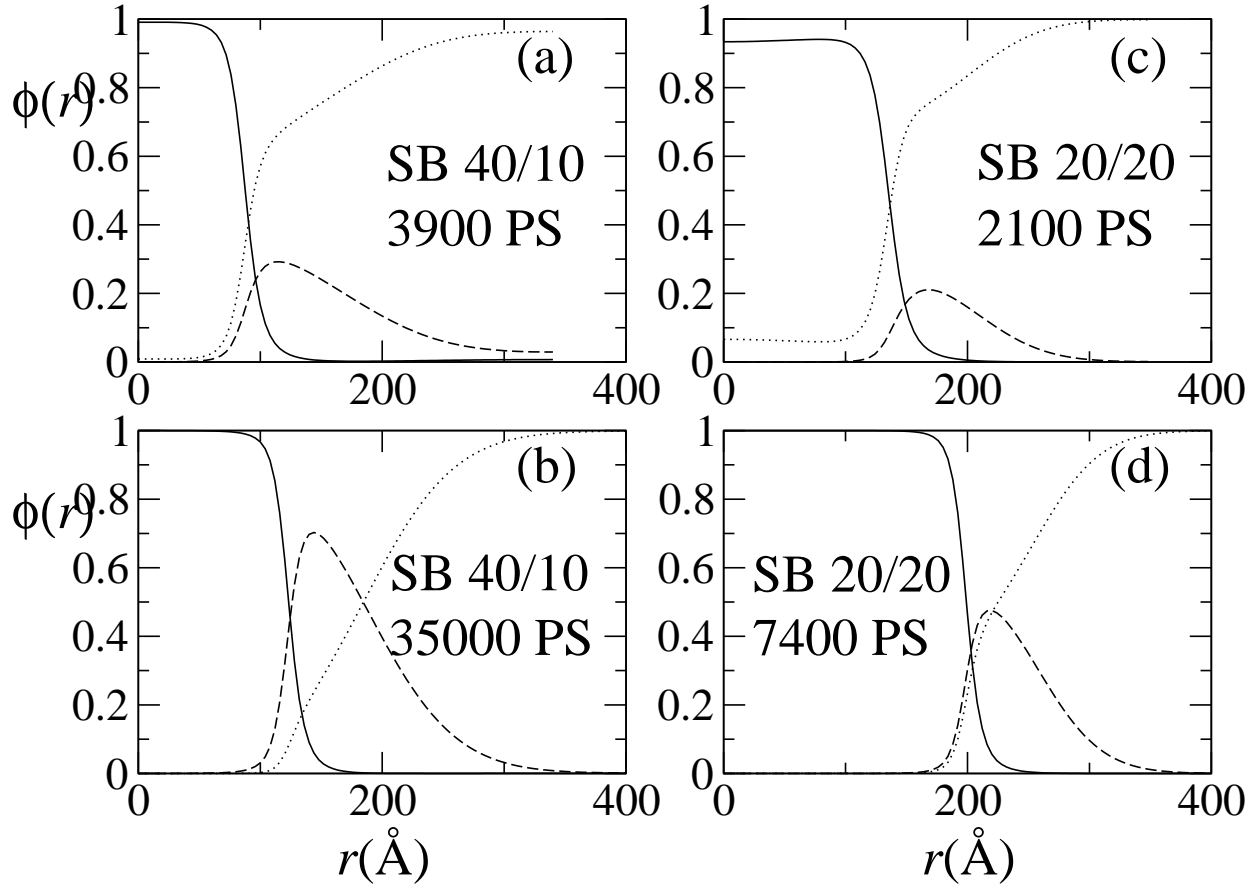}
\caption{\label{cuts-ex} All panels show cuts through the volume
fraction profiles for micelles. The core PB blocks are shown by full
lines, the corona PS blocks by dashed lines, and the polystyrene
homopolymer by dotted lines. Panel (a) shows the profiles for the
asymmetric copolymer SB 40/10 blended with the light homopolymer
3900 PS. Panel (b) shows the corresponding plots for the same
copolymer blended with the very heavy homopolymer 35000 PS. The
decreased mixing of the homopolymer with the corona and
corresponding swelling of the core as the homopolymer molecular
weight is increased may clearly be seen. Panel (c) shows volume
fraction profiles for the strongly solvophobic, symmetric copolymer
SB 20/20 in the light homopolymer 2100 PS, whilst panel (d) shows
the corresponding plots in the heavier 7400 PS. As in panels
(a) and (b), the difference in core size and corona swelling between
the two blends may be clearly seen. In addition, the very light
homopolymer 2100 PS mixes significantly with the core PB blocks in
panel (c). }
\end{figure}

By plotting density profiles of the micelles for the SB 20/20 and SB
40/10 samples blended with the heaviest and lightest homopolymers
considered in the experiments, we may clearly demonstrate the swelling
of the core and
corresponding shrinking of the corona as the homopolymer weight is
increased. All profiles are calculated at $10\,\%$ weight fraction.
In panel (a) of \ref{cuts-ex}, we show cuts through the
density profile of the optimum micelle formed by SB 40/10 in 3900 PS
for the various species. The volume fraction profiles of the core PB
blocks, corona PS blocks and homopolystyrene are shown by full
lines, dashed lines and dotted lines respectively. For this system,
the small homopolymer has a high entropy of mixing with the
polystyrene corona. This means that the corona is strongly swollen
by the homopolymer, and the core shrinks to compensate
\cite{kinning, kinning_winey_thomas}. The opposite extreme may be
seen in panel (b) of \ref{cuts-ex}, where cuts are shown through the
volume fraction profile of a micelle in a blend of SB 40/10
copolymer with the heavy homopolymer 35000PS. Here, the large
homopolystyrene molecules mix much less well with the corona PS
blocks, and the volume fraction of homopolymer in the corona region
is much lower. This results in a clear increase of the core radius
from around $95\,\text{\AA}$ to around $125\,\text{\AA}$.

In the second pair of panels (c and d) in \ref{cuts-ex}, we consider
the symmetric and more strongly solvophobic copolymer SB 20/20.
Again, a decrease in the swelling of the corona by homopolymer and
an increase in the core radius are seen as the sample is blended
with heavier homopolystyrene molecules. In addition, panel (c) shows
clear penetration of the very light homopolymer 2100 PS into the
micelle core.

\section{Conclusions}\label{conclusions}

We have found that self-consistent field theory gives a good
description of all the experimentally-measured trends of spherical
micelle structure. In particular we find that SCFT reproduces the
shape of the variation of $R_c$ with different molecular parameters
much more accurately compared to scaling theory, though like scaling
theory, it overestimates $R_c$ by about 20-30\%. We note however
that our calculations contain no adjustable parameters. Given this
fact, we believe that the agreement between SCFT and experiment for
$R_c$ is good. Since the core radius is the quantity most accurately
measured experimentally, the success of SCFT here is clear
validation of the use of this approach to model micelle formation in
block copolymer/homopolymer blends. For the corona thickness $L_c$,
we find that the accuracy of our SCFT results is at least as good as
those of scaling theory. However we note that this quantity is
measured indirectly \cite{kinning_thomas_fetters} and is subject to
larger experimental error.

SCFT predicts the qualitative trends for the cmc as the homopolymer
molecular weight and copolymer composition and molecular weight are
varied. For smaller core block molecular weights, the predictions
improve on scaling theory by around an order of magnitude. For
heavier core blocks, the agreement between SCFT and experiment
worsens. This problem was attributed to the overestimation by SCFT
of the energy difference between a copolymer in the micelle and one
in the bulk \cite{wijmans_linse}.

In summary, we have shown that SCFT provides a very good
description of micelle structure and hence that it is a suitable
tool for studying the micellization in block copolymer systems. It
is less successful in predicting absolute values of the cmc,
although the correct qualitative trends are obtained. This is
characteristic of a mean-field approximation to a many-body theory:
for example, the mode-coupling theory of the glass transition
\cite{goetze_exp} successfully describes the strength of arrest on
different lengthscales within a structural glass, but is less
effective in predicting the temperature at which this effect occurs
\cite{goetze_exp}.

This work was supported by the UK Technology Strategy Board.

\appendix

\section{Simple Two State Model for Micelle Formation}

Consider a copolymer solution with volume $V$ and copolymer volume
fraction $\phi$ containing a single micelle. To simplify our
calculation, we assume a simple two state model for this system
where there is a sharp demarkation between micelle and bulk so that
copolymer chains can exist either in the micelle or in the bulk and
let the number of copolymer chains in the micelle and bulk be $n$,
$m$ respectively. The free energy of this single micelle system is
then given by\cite{safran_book}
\begin{equation}\label{A}
    A=n f(n)+m k_B T \left(\ln\frac{m}{V}-1 \right)+m f(1)
\end{equation}
where $f(n)$ is the energy per chain for a micelle with aggregation
number $n$ arising from contributions other than the translational
entropy of the copolymer chains. The first term on the R.H.S. of
Equation \ref{A} is the contribution to the free energy from
copolymers in the micelle (neglecting the translational entropy of
the micelle) while the second and third term are due to
translational entropy and non-translational entropy contributions
respectively of copolymers in the bulk. We now assume that $V$ is
the volume per micelle in a many micelle system with total volume
$V_T$ and copolymer volume fraction $\phi$. If we neglect the
translational entropy of micelles, inter-micellar interactions and
micelle polydispersity, the total free energy of the many micelle
system is given by
\begin{equation}\label{Ftot}
    F=\frac{V_T}{V}A.
\end{equation}

To find the equilibrium state of the system, we need to minimize $F$
with respect to $n$, $m$ and $V$ (the internal degrees of freedom of
the system) subject to the constraint that the total number of
copolymer chains is fixed, i.e.,
\begin{equation}\label{constr}
    \frac{V_T}{V}(m+n)=\frac{V_T \phi}{V_0}
\end{equation}
where $V_0$ is the volume per copolymer chain. This is equivalent to
minimizing the modified free energy
\begin{equation}\label{Ftot2}
    F_\mu=\frac{V_T}{V}A-\mu \frac{V_T}{V}(m+n)
\end{equation}
with respect to $m$, $n$ and $V$ where $\mu$ is a Lagrange
multiplier imposing the mass conservation constraint Equation
\ref{constr}.

From $\partial F_\mu/\partial n=0$ we find
\begin{equation}\label{dfdn}
    \frac{\partial A}{\partial n}=f(n)+n f'(n)=\mu
\end{equation}
where $f'(n)$ is the derivative of $f(n)$ with respect to $n$. From
$\partial F_\mu/\partial m=0$ we find
\begin{equation}\label{dfdm}
    \frac{\partial A}{\partial m}=k_B T
    \ln\frac{m}{V}+f(1)=\mu.
\end{equation}
Equations \ref{dfdn} and \ref{dfdm} allow us to identify $\mu$ as
the chemical potential of copolymer chains in the micelle and the
bulk respectively and also tell us that at equilibrium the two
chemical potentials are equal.

From $\partial F_\mu/\partial V=0$ we find
\begin{equation}\label{dfdv}
    A-\mu (m+n)+m k_B T=0.
\end{equation}
Using Equations \ref{A} and \ref{dfdm}, the above equation
simplifies to
\begin{equation}\label{micelle1}
    f(n)=\mu.
\end{equation}
Finally inserting Equation \ref{micelle1} into Equation \ref{dfdn},
we find
\begin{equation}\label{micelle2}
    f'(n)=0.
\end{equation}

Thus minimizing $F_\mu$ with respect to $n$, $m$ and $V$ yields the
equilibrium condition for micellization condition to be that the
free energy per chain in the micelle is minimized (Equation
\ref{micelle2} and the chemical potential of copolymer chains in the
system is equal to the minimized free energy per chain (Equation
\ref{micelle1}. These results agree with the equilibration condition
derived from simple theories for micellization \cite{safran_book},
thus validating the approach used in this paper.

\bibliographystyle{apsrev}

\end{document}